\documentclass[10pt,conference]{IEEEtran}
\IEEEoverridecommandlockouts
\usepackage{cite}
\usepackage{csquotes}
\usepackage{amsmath,amssymb,amsfonts}
\usepackage{algorithmic}
\usepackage{graphicx}
\usepackage{textcomp}
\usepackage{float}
\usepackage{makecell}

\usepackage{xcolor}
\def\BibTeX{{\rm B\kern-.05em{\sc i\kern-.025em b}\kern-.08em
    T\kern-.1667em\lower.7ex\hbox{E}\kern-.125emX}}
\begin{document}

\title{LLM-Powered Quantum Code Transpilation}



\author{\IEEEauthorblockN{Nazanin Siavash}
\IEEEauthorblockA{Department of Computer Science \\
University of Colorado Colorado Springs (UCCS)\\
United States \\
nsiavash@uccs.edu}
\and
\IEEEauthorblockN{Armin Moin}
\IEEEauthorblockA{Department of Computer Science \\
University of Colorado Colorado Springs (UCCS)\\
United States \\
amoin@uccs.edu}
}

\maketitle


\begin{abstract}
There exist various Software Development Kits (SDKs) tailored to different quantum computing platforms. These are known as Quantum SDKs (QSDKs). Examples include but are not limited to Qiskit, Cirq, and PennyLane. However, this diversity presents significant challenges for interoperability and cross-platform development of hybrid quantum-classical software systems. Traditional rule-based transpilers for translating code between QSDKs are time-consuming to design and maintain, requiring deep expertise and rigid mappings in the source and destination code. In this study, we explore the use of Large Language Models (LLMs) as a flexible and automated solution. Leveraging their pretrained knowledge and contextual reasoning capabilities, we position LLMs as programming language-agnostic transpilers capable of converting quantum programs from one QSDK to another while preserving functional equivalence. Our approach eliminates the need for manually defined transformation rules and offers a scalable solution to quantum software portability. This work represents a step toward enabling intelligent, general-purpose transpilation in the quantum computing ecosystem.

\end{abstract}


\section{Introduction}
Large Language Models (LLMs) trained on a vast amount of source code and technical documentation have demonstrated remarkable capabilities in various software engineering tasks \cite{Bhatia+2024,Yang+2024,Hagos+2024}. These models bypass the need for domain-specific retraining, leveraging their pre-trained knowledge to reason about code structure, semantics, and intent. They can also generate code in the desired target language effectively and efficiently. As such, they offer a promising alternative to traditional rule-based transpilers, which require extensive manually defined rules and deep domain expertise \cite{Roziere+2020}.

While LLMs have shown utility in classical software development, their potential in quantum programming, and specifically quantum code transpilation (i.e., translation), remains largely unexplored. Quantum software development presents unique challenges due to the heterogeneity of the platforms and the novel programming and computation models based on the unique characteristics of quantum computers. Transpiling quantum code from one QSDK (e.g., IBM's Qiskit) to another (e.g., Google's Cirq) is critical for achieving cross-platform interoperability and hardware abstraction in quantum computing workflows.

The contribution of this extended abstract accompanying our poster is to propose leveraging LLMs as quantum code transpilers. In what follows, we briefly review the literature and propose our novel approach. We then present our preliminary findings, conclude, and suggest future work.



\section{Related Work}
Stefano et al. \cite{Stefano+2024} examined how transpiling quantum circuits to various gate sets affects the emergence of quantum-specific code smells. Their findings suggested that the choice of gate set plays a significant role in shaping the structural characteristics of the resulting circuits, which in turn influence the presence and severity of code smells. The study particularly emphasized that variations in gate-level abstraction and optimization techniques lead to structural differences, a pattern that was especially evident in experiments conducted on synthetic benchmarks. The quantum programming language used for the original circuits was Qiskit, and the transpilation process involved converting these Qiskit-based circuits into different gate sets for execution on various quantum hardware platforms. In another study, Arulandu \cite{ARULANDU2024} developed a Python-based transpiler designed to convert OpenQASM 3.0 programs into CUDA-Q kernels. Building on existing OpenQASM parsing tools, the transpiler supports a substantial subset of the OpenQASM 3.0 specification, including features such as custom gates, control and adjoint modifiers, and binary expressions. The system was validated through a combination of custom unit tests and randomized evaluations on Clifford circuits with varying sizes and depths. Furthermore, the development process contributed to enhancements in open-source tooling for circuit optimization, semantic analysis, and program unrolling, thereby advancing the broader ecosystem for quantum programming.

\section{Proposed Approach}
In this paper, we propose a framework-oriented perspective for the transpilation of quantum code. While there has been notable progress in the transpilation of classical code using LLMs, targeting quantum code transpilation via LLMs remains largely unexplored. Our methodology positions an LLM as a \textit{code transpiler} capable of converting quantum programs written in one QSDK (referred to as QSDK 1) into functionally equivalent code in another QSDK (namely, QSDK 2). This transpilation process enables interoperability between heterogeneous quantum programming environments (e.g., Qiskit to Cirq and vice versa), reducing the manual effort required to port quantum programs across QSDKs.


The workflow for the proposed quantum code transpilation consists of the following key steps: (i) Input Specification: The quantum program, originally written in QSDK 1, is parsed and annotated to serve as input for the LLM. This step ensures that all relevant constructs are clearly defined and formatted for downstream processing. (ii) Prompting: A domain-specific prompt is designed to guide the LLM in understanding the syntactic and semantic structures of both the source (QSDK 1) and target (QSDK 2) languages. Potential prompts could include: \textit{a) Task Description:} A concise instruction defining the transpilation objective (e.g., ``Convert the following Qiskit code into its equivalent in Cirq.'') \textit{b) Mapping Instructions:} Explicit rules or examples outlining how specific constructs in \textit{QSDK 1} map to those in \textit{QSDK 2} (e.g., gate translations and qubit initialization). (iii) Code Transpilation: The LLM generates the corresponding code in \textit{QSDK 2}.


\section{Preliminary Findings}
In our preliminary experiments, we employ LLaMA 3 (8B), GPT-2, and GPT-4o as LLM-based transpilers using both zero-shot and one-shot prompting strategies. The objective is to transpile quantum programs written in Qiskit into equivalent programs using Cirq, maintaining functional equivalence while conforming to Cirq’s syntax and structure. The provided Qiskit code constructs a two-qubit quantum circuit with two classical bits. It performs the following operations: (i) It applies a Hadamard gate to qubit 0, placing it in superposition. (ii) It applies a CNOT gate between qubits 0 (control) and 1 (target), entangling them to form a Bell state. (iii) It measures both qubits and stores the results in classical bits.

To assess the quality of the generated Cirq code, we define a custom Code Quality Metric that rewards using some expected patterns, which are predefined Cirq-specific constructs, such as circuit creation, qubit initialization, gate operations, measurements, and simulation. We then apply the penalties of 0.05 per compiler warning (e.g., unused imports) and 0.2 per compiler error, constrained to the score range of [0.0, 1.0]. For LLaMa 3, we begin with a generic prompt in the zero-shot prompting setting, which successfully generates Cirq code but results in non-compilable output. We subsequently employ GPT-4o to analyze the faulty transpilation and offer corrective suggestions. Based on this feedback, we refined the prompt to include stricter constraints, significantly improving the output's correctness and syntax compliance. In the one-shot prompting setting, we provide the model with an explicit example of Qiskit-to-Cirq transpilation. This guides the LLM toward more accurate and structurally valid Cirq code, enhancing consistency and reducing post-processing effort. While GPT-2 fails to produce meaningful transpilation results in both zero-shot and one-shot prompting scenarios, GPT-4o yields varying code quality metrics across runs. This variability highlights GPT-4o’s high degree of creativity, as it is capable of generating multiple diverse and valid outputs for the same task prompt. All transpiled outputs are evaluated using our proposed code quality metric, allowing us to quantitatively compare outputs across models and prompting strategies. The preliminary results for each transpiler are presented in Table~\ref{llm_code_quality}.

\begin{table}[h]
\centering
\caption{Code quality analysis for various LLMs under different prompting strategies}
\begin{tabular}{|>{\centering\arraybackslash}m{1.2cm}|>{\centering\arraybackslash}m{1.3cm}|>{\centering\arraybackslash}m{1cm}|>{\centering\arraybackslash}m{3.8cm}|}
\hline
\textbf{LLM} & \textbf{\# Shots} & \textbf{Code Quality Value} & \textbf{Low-Performance Explanation} \\
\hline
LLaMa 3 & Zero-Shot & 0.78--1 & Missing expected Cirq pattern: Measurement operations \\
\cline{2-4}
& One-Shot & 1 & -- \\
\hline
GPT 2 & Zero-Shot & \textit{Failed} & Lack of domain-specific knowledge \\
\cline{2-4}
& One-Shot & \textit{Failed} & Lack of domain-specific knowledge \\
\hline
GPT-4o & Zero-Shot & 0.52--1 & Missing expected Cirq pattern: Basic gates, Measurement operations and Unused imports\\
\cline{2-4}
& One-Shot & 1 & -- \\
\hline
\end{tabular}
\label{llm_code_quality}
\end{table}

\section{Conclusion and Future Work}
This work has introduced a novel approach to automated code transpilation for quantum platforms, focusing on the IBM Qiskit and the Google Cirq. As part of our future work, we plan to improve the contextual accuracy and reliability of the transpiled code by integrating a Retrieval-Augmented Generation (RAG) pipeline. We also plan to employ well-established metrics, such as Transpilation Coverage \cite{ShiraishiShinagawa2024}, to precisely evaluate the quality and correctness of the code transpiled by LLMs. Furthermore, we plan to extend this transpilation to support advanced quantum features.
\section*{Acknowledgment}
This work is funded by a grant (Q-Dev) from the Colorado Office of Economic Development and International Trade (OEDIT).

\bibliographystyle{ieeetr}
\bibliography{references}

\end{document}